# Systematic Review of Cybersecurity in Banking: Evolution from Pre-Industry 4.0 to Post-Industry 4.0 in Artificial Intelligence, Blockchain, Policies and Practice

Tue Nhi Tran[1]

[1] The University of Sheffield, Sheffield S10 2TN, UK

tntran2@sheffield.ac.uk

**Abstract.** Throughout the history from pre-industry 4.0 to post-industry 4.0, cybersecurity at banks has undergone significant changes. Pre-industry 4.0 cyber security at banks relied on individual security methods that were highly manual and had low accuracy. When moving to post-industry 4.0, cybersecurity at banks had a major turning point with security methods that combined different technologies such as Artificial Intelligence (AI), Blockchain, IoT, automating necessary processes and significantly increasing the defence layer for banks. However, along with the development of new technologies, the current challenge of cybersecurity at banks lies in scalability, high costs and resources in both money and time for R&D of defence methods along with the threat of high-tech cybercriminals growing and expanding. This report goes from introducing the importance of cybersecurity at banks, analyzing their management, operational and business objectives, evaluating pre-industry 4.0 technologies used for cybersecurity at banks to assessing post-industry 4.0 technologies focusing on Artificial Intelligence and Blockchain, discussing current policies and practices and ending with discussing key advantages and challenges for 4.0 technologies and recommendations for further developing cybersecurity at banks.

**Keywords:** cybersecurity, banking, Artificial Intelligence, Blockchain, financial security, information system management.

# 1 Introduction

Banks play a crucial role in managing the flow of funds between savers and borrowers, including individuals, businesses, and governments [1]. Despite being the largest profitgenerating sector globally, there is skepticism about its long-term value creation, with banking ranking lowest among sectors based on price-to-book ratios [2].

Among many challenges banks face, cybersecurity risk is identified as the leading threat to growth in the banking sector, surpassing all other risks, according to the 2024 U.S. Banking Industry Outlook Survey [3]. The financial impact of cybersecurity incidents on banks is significant, with the global average annual cost of cybercrime for banks rising to approximately $18.5 million, the highest among all industries [4].

This issue is further highlighted by past crises, such as the high-profile attacks during 2015 and 2016, which led to substantial losses, including RUB 140 million for Russian bank, nearly $2.2 million from First Bank ATMs, and $81 million stolen from the Central Bank of Bangladesh [5]. In response to these risks, 55% of executives are increasing their budgets to address cyber threats, despite the belief that many banks already have adequate cybersecurity measures in place [6].

This paper explores how cybersecurity in banks has evolved, examining the use of technologies before and after Industry 4.0, identifying key management, operational, and business goals, and providing insights on benefits and challenges of these technologies along with recommendations for executives to enhance cybersecurity at banks.

The objectives of this paper are as follows:

1. To identify key management, operational and business objectives of cybersecurity at banks.

2. To review pre-industry 4.0 technologies and post-industry 4.0 technologies in Artificial Intelligence and Blockchain for cyber defence at banking.

3. To critique current policies and practices for financial cybersecurity.

4. To present current obstacles and opportunities for AI and Blockchain for financial cybersecurity, and give recommendations for banks.

The organization of this work is as outlined below. Section two provides sector's objectives, section three synthesise pre-industry 4.0 technologies for cyber security. Then section four continues to summarise AI and Blockchain technologies for cybersecurity at banks. Section five discusses current policies and practices, before concluding the paper in section six with discussion, recommendation and conclusion for the report.

## 2 Cybersecurity's Objectives at Banks

This section will explore the management, operational, and business objectives of cybersecurity in banks using Porter's value chain analysis [7] (refer Figure 1).

Management objectives in cybersecurity for banks focus on technology development, effective human resource management, and building firm infrastructure. According to the NIST Cybersecurity Framework, a key part of technology development involves integrating secure software practices throughout the software lifecycle to address risks, particularly those posed by new technologies [8]. In human resource management, access control and identity management systems are essential for restricting access to authorized personnel, ensuring compliance with security policies [9]. Additionally, infrastructure objectives involve integrating security architecture with the bank's risk strategy to protect assets, ensuring confidentiality, integrity, and availability while creating resilient systems [10].

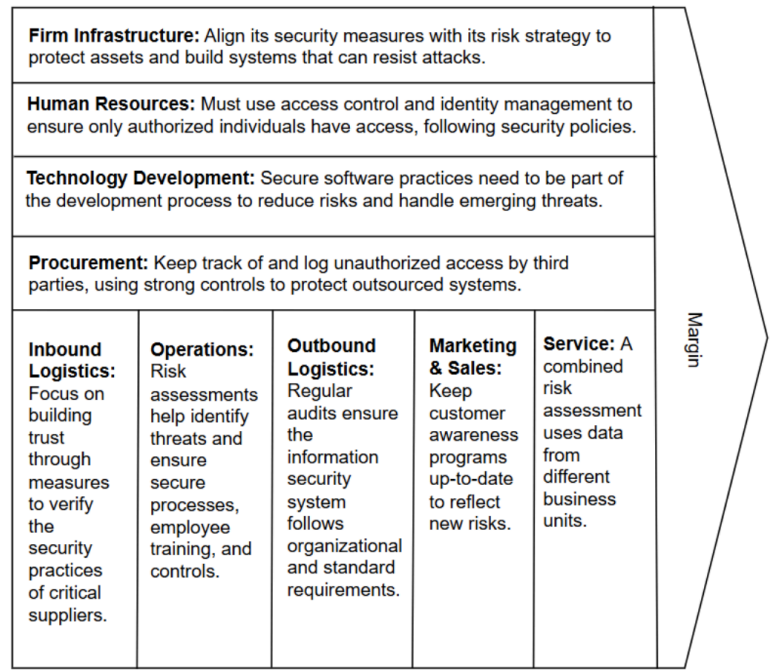

**Fig. 1.** Porter's Value Chain Analysis for Banks' Objectives for Cybersecurity

The operational objectives of cybersecurity in banks address critical areas, including inbound logistics, operations, outbound logistics, and procurement, to establish a strong security framework. In inbound logistics, according to the National Institute of Standards and Technology in the United States, banks prioritize confidence-building measures, such as third-party assessments, on-site visits, and formal certifications, to evaluate and ensure the security practices of their critical suppliers. For operations, following the Federal Financial Institutions Examination Council in the United States, a comprehensive risk assessment helps identify emerging threats, secure processes, enhance employee training, and implement effective controls for vital areas like customer call centers and IT help desks. In outbound logistics, regular internal audits ensure that the information security management system complies with organizational requirements and international standards [10]. Lastly, in procurement, banks focus on monitoring unauthorized access by third-party providers and identifying all users accessing systems to ensure access rights align with the functionality, criticality, and risks associated with financial institution systems and data [11].

The business objectives of cybersecurity in banks focus on key areas like marketing & sales and services to ensure secure and trustworthy operations. In marketing and sales, banks aim to keep customer awareness programs updated to reflect new risks, such as those introduced by faster payment services, while ensuring that marketing materials align with the communication of security risks to avoid potential compliance issues [11]. Accordingly, for services, banks strive to implement an integrated risk assessment approach that gathers input from fraud research, customer service, and cybersecurity teams, along with reports of attempted or actual fraud from customers, to identify and mitigate emerging authentication threats effectively.

## 3 Methodology

This study follows a systematic review approach to analyze the evolution of cybersecurity in banking, comparing developments from Pre-Industry 4.0 (1990–2010) to Post-Industry 4.0 (2011–2025). The research focuses on cybersecurity strategies, the role of Artificial Intelligence and Blockchain, and the effectiveness of various security frameworks in banking. A structured methodology was adopted to ensure a rigorous and transparent approach to data collection, study selection, and synthesis.

### 3.1 Eligibility Criteria

The selection of studies was guided by predefined inclusion and exclusion criteria to ensure relevance and academic rigor. Only peer-reviewed journal articles, conference papers, and research publications directly addressing cybersecurity in banking were included. Studies that explored cybersecurity risk mitigation strategies, security frameworks and technological applications in banking were prioritized. Additionally, separate searches were conducted for AI and Blockchain to examine their specific roles in banking cybersecurity in the Post-Industry 4.0 era.

Studies were excluded if they lacked empirical data, did not focus on banking cybersecurity, or were non-peer-reviewed sources such as white papers, patents, books and industry reports. Research that provided only a broad discussion of cybersecurity without application to banking was also omitted. The selected studies were categorized into three primary themes: general cybersecurity in banking (Pre-Industry 4.0 and Post-Industry 4.0), AI-driven cybersecurity approaches in banking (Post-Industry 4.0), and Blockchain applications in banking cybersecurity (Post-Industry 4.0).

### 3.2 Search Strategy

The review was conducted using four leading academic databases: IEEE Xplore, ACM Digital Library, SpringerLink, and Scopus. These databases were chosen for their extensive coverage of research related to cybersecurity, financial technology, and banking security.

A structured search strategy was implemented, applying different search terms based on the research focus. For Pre-Industry 4.0 cybersecurity trends, the search query "cybersecurity AND (bank OR banking)" was applied across IEEE Xplore, ACM, Springer, and Scopus.

For Post-Industry 4.0, the same general query "cybersecurity AND (bank OR banking)" was used to analyze the overall publication trends in banking cybersecurity research. When it comes to examine AI's role in banking cybersecurity during this period, the query was refined to "cybersecurity AND (bank OR banking) AND (AI OR Artificial Intelligence)". Similarly, to analyze Blockchain applications in banking security, the search term "cybersecurity AND (bank OR banking) AND blockchain" was applied. This approach ensured a focused and systematic assessment of how AI and Blockchain have been integrated into banking cybersecurity.

### 3.3 Selection Process

The selection of studies followed a two-stage screening process. First, article titles and abstracts were screened by a reviewer to assess relevance. In the second stage, fulltext articles were examined to ensure they met the inclusion criteria. Studies that did not focus specifically on cybersecurity in banking or lacked methodological rigor were excluded. The screening process was conducted manually to maintain accuracy and minimize potential selection bias.

### 3.4 Data Collection Process

Data were extracted systematically from each selected study to ensure a structured analysis. Extracted information included publication details (title, authors, year, and source), cybersecurity focus (general banking security, AI-driven security, or Blockchain security), technological period (Pre-Industry 4.0 or Post-Industry 4.0), and findings related to effectiveness and challenges in cybersecurity implementation.

The primary outcome of data collection process was the effectiveness of cybersecurity methods in banking, assessed through their impact on risk mitigation, fraud prevention, and financial data protection. Secondary outcomes included the adoption trends, benefits and challenges of AI and Blockchain in banking security.

### 3.5 Synthesis Method

A narrative synthesis was conducted to compare findings across different research themes. Studies were grouped into Pre-Industry 4.0 and Post-Industry 4.0 cybersecurity approaches with AI-driven security strategies and Blockchain-based financial security frameworks. The findings were analyzed to identify patterns, emerging trends, and gaps in cybersecurity research.

Trend analysis graphs were used to visually represent the growth of cybersecurity research in banking, allowing for a clearer understanding of how the industry has evolved over time. This comparative approach facilitated an in-depth examination of how AI and Blockchain are shaping the future of banking security.

Additionally, to provide a deeper evaluation of the impact of AI and Blockchain on cybersecurity in banking, Leavitt's Diamond Model was applied. This framework considers four interconnected components including tasks, people, technology, and structure, to assess how AI and Blockchain influence security practices in banking institutions.

### 3.6 Quality Assessment

To minimize reporting bias, multiple academic databases were searched, and only peerreviewed sources were included. The study ensured diversity by selecting research from different regions and institutions, preventing an overrepresentation of specific markets. Additionally, by excluding industry white papers and commercial reports, the study maintained a high level of academic rigor.

The effectiveness of cybersecurity strategies was assessed by analyzing publication frequency, case studies, and reported success rates in reducing cybersecurity risks. For general cybersecurity trends, the study measured the number of publications over time to track research growth in banking security. For AI and Blockchain, a more detailed evaluation was conducted to determine their impact on fraud detection, transaction security, and financial risk management.

## 4 Pre-Industry 4.0 Cybersecurity Methods at Banks

The following section outlines technologies used in pre-industry 4.0 for protecting security at banks along with assessing their popularity in academic databases, then critically evaluate each type of security.

### 4.1 The Popularity of Cybersecurity Methods at Banks in Pre-Industry 4.0 Era

Before Industry 4.0, cybersecurity in banks was not widely researched, with limited publications in the 1990s across four major academic databases including IEEE Xplore, ACM, Springer and Scopus.

The gradual increase in publications of cyberscurity at banks since 2001 with the sudden surge in the IEEE Xplore with almost 40 research papers reflects the growing recognition of the importance of cybersecurity in the banking sector as digital transformations took hold (refer Figure 2).

### 4.2 Critical Evaluation of Cybersecurity Methods at Banks in Pre-Industry 4.0

Pre-industry 4.0 methods for cybersecurity at banks were foundational for their time but were largely manual and reactive, reflecting the technological limitations of the era (refer Table 1).

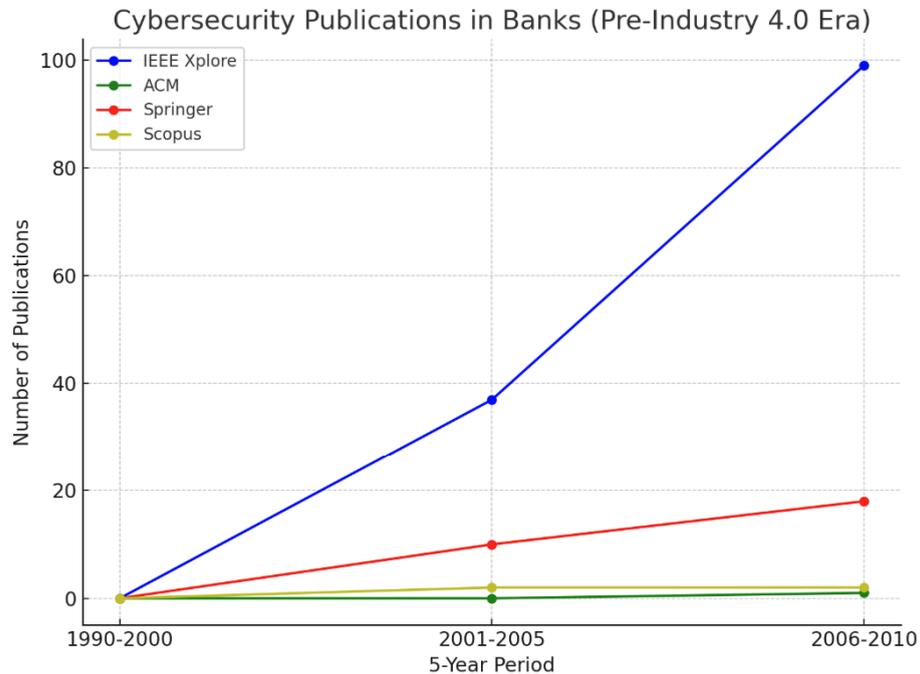

**Fig. 2.** Trends in Cybersecurity Publications for Banks Pre-Industry 4.0 (1990–2010) [13]

Access control mechanisms are widely used in banks to restrict access to authorized personnel, which helps meet management objectives like reducing insider threats and maintaining secure systems [12]. They also play a key role in operational objectives, such as securing supply chains and sensitive information. Nevertheless, security policies to control the access are not fully comprehensive, the implementation methodology is not clear, along with the vulnerabilities in the bank with potential data leaks make the access control practice in the banks alone still put the banks at risk.

Education and training programs help banks meet management objectives by addressing human vulnerabilities and improving employee awareness [14, 15]. They also support operational goals by reducing errors caused by unintentional human actions. Yet, inconsistent delivery and gaps in customer knowledge limit their impact. To make training more effective, banks should focus on delivering tailored and standardized programs, which would also enhance customer trust as part of their business goals.

Threat modeling is becoming more recognized for its ability to identify vulnerabilities during the system design phase. This method helps banks meet management objectives by improving technology development and supports operational goals by enhancing risk assessments [16, 17]. However, its reliance on predefined scenarios can result in overlooked risks. Introducing more adaptive tools to identify emerging threats could make threat modeling more effective and align it better with business objectives, such as building customer confidence and reducing fraud.

Table 1. Cybersecurity Methods in Banking Before Industry 4.0 (1990–2010).

| Cybersecurity Method | References | Strengths | Weaknesses |
|---|---|---|---|
| Access Control Mechanisms | [12] | Improves policy framing through awareness. Reduces insider threats with consistent employee training. | Strong reliance on access controls. Limited employee expertise in technical areas. |
| Education and Training | [14, 15] | Raises awareness among customers. Reduces errors caused by human behavior through training programs. | Gaps in customer knowledge may increase risks. Inconsistent policy implementation across institutions. |
| Threat Modeling | [16, 17] | Helps uncover vulnerabilities during system design. Extended threat models improve accuracy in risk assessment. | Limited to predefined scenarios, which may leave certain risks unaddressed. |
| One-Time Password (OTP) Systems | [18, 19] | Provides robust, session-specific security by using unique fingerprint for generating OTP to create a different password key for each session | Predictability issues when fingerprint lacks variation. Usability can decline with frequent updates. |
| Digital Signature Mechanisms | [20, 21] | Minimizes reliance on potentially compromised devices. Validates transactions efficiently, reducing user errors | Dependent on secure handling of devices. Commercial solutions can incur high maintenance costs. |
| Encryption Technologies | [22–24] | Provides strong protection for | Poor logging practices can hinder auditing and evaluation. |

| | | data in transit and at rest. Dual signature methods enhance data integrity. | Static encryption methods may lead to vulnerabilities over time. |
|---|---|---|---|
| Defense-in-Depth | [25] | Reduces the likelihood of successful threats with layered security measures. Proactive monitoring enhances responsiveness. | Costly to implement and maintain. Additional layers might not always yield proportional benefits. |
| Anti-Phishing Techniques | [15, 26] | Effectively combats unauthorized access. | Relies on users remembering anti-phishing codes, which may not always be practical. |

OTP systems are a highly popular method for authentication due to their ability to provide session-specific security [18, 19]. Their use of biometrics adds a layer of uniqueness, helping meet management goals related to secure technology. They also support operational objectives by protecting transactions. The downsides of OTP systems are the duplication of OTP with the same fingerprint and threats of password stealing, which might further cause dangers for customers being exploited by cybercriminals.

The digital signature mechanisms help to prevent man-in-the-middle attacks, which reduces the mistakes of misidentifying customers and aligns with management's objectives in security policies [20, 21]. This methodology also ensures the integrity of data via visual assessment when doing transactions to support operational and business goals. The only concerns here is the reliability of the device handling this method and the maintenance cost for the universal use. The future of implementing public key cryptography signature technology also brings a lot of hope in public verification for any transaction to increase customer trust.

Encryption could be said that one of the most popular methods for cybersecurity in banks for its usefulness in communication and storage security [22]. This method matches the management objectives in being a secure protocol for the online banking system, serves as a guaranteed e-payment method for operations and protect the confidential data for customers in business [23, 24]. However, any overlook on any part of key management in encryption could cause severe threatens to the bank's security, which is a severe concern if the banks only depend on encryption as the only defence layer.

Defense-in-depth strategies are highly effective for layered security to meet management objectives by building resilient infrastructures and support operational goals through proactive monitoring [25]. However, simplifying these strategies and optimizing resources could make them more accessible and effective for achieving business objectives like improving overall security and efficiency.

Anti-phishing techniques are effective but not as commonly used due to their reliance on user memory and behavior [15, 26]. They support management objectives by reducing unauthorized access and operational goals by protecting customer information. To increase their adoption, banks need to improve the usability of these techniques and make them less reliant on users, which would also help align them better with business objectives like fostering trust and improving customer experience.

## 5  Post-Industry 4.0 Cybersecurity Methods at Banks

This section focuses on the evolution of cybersecurity techniques in banks after the period of Industry 4.0 and how the banks are gradually relying on smart technologies such as the use of AI and blockchain to overcome the broader threatening cyberattacks.

### 5.1  The Popularity of Cybersecurity Methods at Banks in Post-Industry 4.0 Era

The post-industry 4.0 era has seen a notable rise in cybersecurity research focused on banking since 2016 with the peak at Springer with over 500 research papers (refer Figure 3).

The increasing research volume highlights the growing focus on addressing advanced cybersecurity threats in banking. The largest surge occurred between 2016–2020 and 2021–2024, emphasizing the urgent need for innovative solutions to safeguard financial systems.

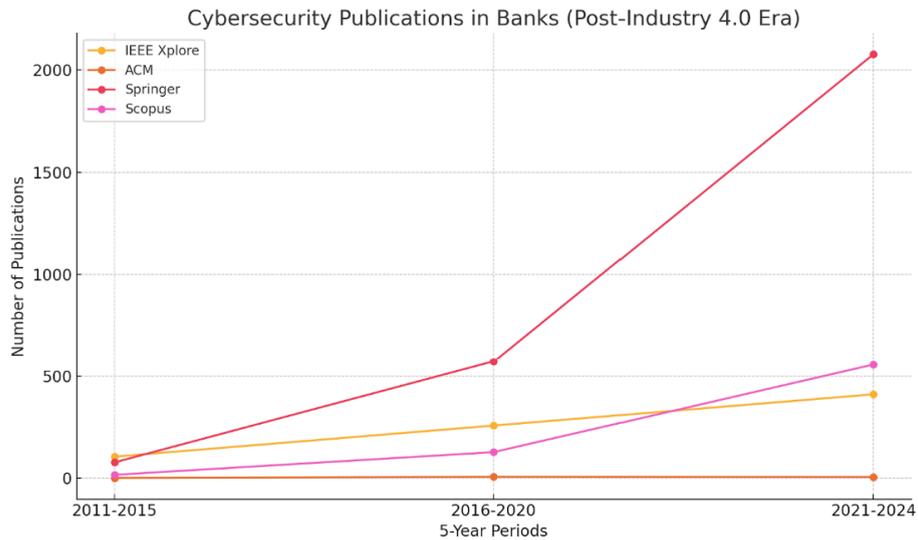

**Fig. 3.** Trends in Cybersecurity Publications for Banks Post-Industry 4.0 (2011–2024) [27]

### 5.2 Critical Evaluation of Cybersecurity Methods at Banks in Post-Industry 4.0

This section assesses the efficiency of cybersecurity measures at the banks post-industry 4.0 era for the transformable roles of AI and blockchain in ensuring security, risk management, and fraud prevention.

#### 5.2.1 Artificial Intelligence

##### 5.2.1.1 Critical Evaluation of Artificial Intelligence Methods Used in Cybersecurity at Banks

For AI cybersecurity at banks in post-industry 4.0, management objectives are supported by advanced detection and prediction, though scalability and complexity pose challenges (refer Table 2). Operational goals benefit from real-time detection but face data quality limitations. Business objectives improve with stronger security, though biases and costs hinder consistent service delivery.

**Table 2.** Artificial Intelligence Cybersecurity Methods in Banking After Industry 4.0 (2011-2024).

| Cybersecurity Method | References | Strengths | Weaknesses |
|---|---|---|---|
| Artificial Neural Networks (ANN) | [28] | High accuracy in intrusion detection and low false positive rate. Real-time analysis across multiple machines. | Risk of overfitting due to complex models. Difficulty handling increasing network traffic in real-time. |
| Defensive, Offensive, and Adversarial AI | [29] | Improves detection accuracy. Simulates potential cyberattacks, and strengthens system defenses through testing. | Misuse of offensive techniques is a concern. Both testing and defenses rely on in-depth system knowledge. |
| Quantum Cryptographic Techniques with AI | [31] | Enhances encryption security and swiftly identifies system vulnerabilities. | Implementation is highly complex. Challenging to align with existing encryption systems. |
| Machine Learning, Deep Learning and Blockchain Intelligence | [30] | Excels at identifying threats, automating security processes, and adapting to emerging risks. | Susceptible to adversarial manipulation. Biased training data and difficulties in scaling for large datasets. |
| Multi-Task Learning (MTL) | [32] | Efficiently manages multiple security issues. Particularly effective in resourceconstrained environments. | Needs large datasets for effective training. Requires careful design and tuning for efficiency. |

With high accuracy and low false positive rate, ANN supports management objectives by enabling proactive risk management and operational goals through realtime threat detection [28]. It strengthens business objectives by enhancing customer trust. However, risks like overfitting and challenges with increasing traffic may limit long-term effectiveness and harm reputation.

AI Defensive and Offensive Cybersecurity Strategies align with management objectives by enhancing strategic defense and operational goals through large-scale anomaly detection [29]. Business objectives benefit from stronger customer-facing security, but the dual-edged nature of these methods and limited focus on human factors pose risks, including reputational harm from misuse.

Machine Learning, Deep Learning and Blockchain Intelligence align with management objectives by strengthening data integrity and operational goals through automated security processes [30]. Business objectives are supported by enhanced customer trust. However, scalability challenges and data integrity risks limit strategic and operational efficiency. AI integrating with Quantum Cryptographic Techniques aids management objectives with predictive analytics and operational goals through robust encryption [31]. Business objectives benefit from enhanced customer trust. However, complexity and high costs hinder strategic adoption, while resource-intensive processes affect operational workflows and scalability.

Multi-Task Learning supports management objectives by enabling multi-task efficiency, operational goals through improved threat handling, and business objectives via robust customer security [32]. Challenges include model complexity and high-quality data requirements, which may limit adaptability and customer protection.

Threat intelligence systems enhance management objectives with proactive planning and operational goals through efficient scenario generation [33]. Business objectives are supported by updated customer awareness programs. Nevertheless, biases and data quality dependencies may limit strategic and operational effectiveness, impacting customer trust.

*5.2.1.2 Analyse Artificial Intelligence Methods Used in Cybersecurity at Banks under Leavitt's Diamond Model*

The integration of advanced AI technologies, including Artificial Neural Networks, Multi-Task Learning, Defensive and Offensive AI and Forecasting, has become a vital component of modern cybersecurity strategies in banking. These tools enhance the ability to detect threats, improve encryption strength, and provide adaptability to address the ever-evolving landscape of cyber risks [28, 29, 32, 33].

Collaboration between human expertise and AI is essential for effective cybersecurity implementation. Banks are emphasizing employee training to ensure seamless integration of AI tools into existing frameworks. Practical exercises addressing industry-specific challenges, combined with collaboration across departments are key to maximizing the potential of these advanced solutions.

AI-driven advancements, such as Machine Learning, Deep Learning, and automation, are reshaping cybersecurity operations. Predictive analytics enables earlier detection of potential threats, while automation reduces human errors and enhances overall efficiency [30].

Post-industry 4.0, AI plays a critical role in real-time monitoring for anomalies, automated responses to threats, and dynamic planning to address emerging risks for banks.

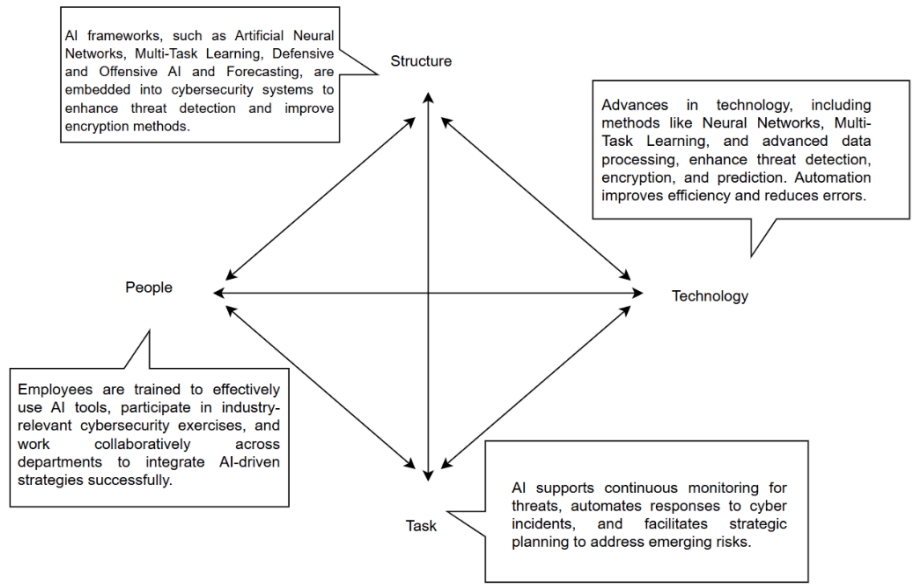

**Fig. 4.** Artificial Intelligence Cybersecurity Methods in Banking under Leavitt's Diamond model

### *5.2.2    Blockchain*

*5.2.2.1    Critical Evaluation of Blockchain Methods Used in Cybersecurity at Banks*

Blockchain technology has revolutionized the banking sector by enhancing security, building trust, and improving operational efficiency (refer Table 3).

Blockchain-based Adaptive Algorithm addresses unauthorized access and distributed denial-of-service attacks, supporting management objectives [34]. Operationally, it ensures uninterrupted service with high robustness and accuracy, fostering reliability and customer trust. On the business front, it enhances reputation and client confidence. However, high implementation costs and integration challenges hinder its widespread adoption.

Distributed Ledger Technology reduces risks associated with centralized systems, improving fraud prevention and transaction integrity while ensuring transparency [35, 36]. It supports business goals by fostering trust through immutable records, though privacy concerns and regulatory hurdles complicate its application.

Table 3. Blockchain Cybersecurity Methods in Banking After Industry 4.0 (2011-2024).

| Cybersecurity Method | References | Strengths | Weaknesses |
|---|---|---|---|
| Blockchainbased Adaptive Neuro-Fuzzy K-Nearest Neighbors Algorithm | [34] | Strong data protection through cryptography. High accuracy, robustness and reliable transaction services. | High hardware costs. Complex integration with existing systems. |
| Distributed Ledger Technology (DLT) | [35, 36] | Decentralized and secure framework with strong consensus mechanisms. Transparent, tamper-proof data. | Greater vulnerabilities due to increased access points and challenging infrastructure changes. Potential privacy risks and scalability constraints. |
| Secure Transaction Platforms and Real-Time Monitoring Systems | [37] | Transparent operations build trust and cost-effective processes. Secure transaction methods prevent tampering. | High energy usage and potential user resistance. Lack of regulatory clarity. |
| Blockchain with Convolutional Neural Network for Biometric Verification | [30] | Enhanced security through decentralization. Tamper-resistant data storage and improved data privacy. | Scalability remains a challenge and requires high technical expertise. Uncertainty in regulatory compliance |
| Decentralized Blockchain | [39] | Lower transaction costs and faster processing of payments. Strong security and integrity through cryptographic techniques. Regulatory hurdles for traditional banks and risk of misuse. | Resource-intensive setup and complex adoption for traditional systems. Struggles with high transaction volumes. |
| Smart Contracts | [40] | Automates transactions securely. Reduces operational costs and promotes financial inclusion. | Regulatory hurdles for traditional banks and risk of misuse. Legal uncertainties for automated contracts. |
| Blockchain Integrated with Federated Learning | [41] | Decentralized and secure data sharing along with immutability ensuring data trustworthiness. Privacy-preserving federated learning and traceability aids compliance. | Complex implementation process and scalability issues with large transaction volumes. Evolving regulatory landscape complicates adoption. |

Secure Transaction Platforms and Real-Time Monitoring Systems align with management goals by enabling effective fraud detection and oversight [37]. They enhance resilience and reduce operational costs while supporting business objectives through increased customer trust. However, technical limitations and user resistance pose challenges.

Biometric Verification Systems integrated with Blockchain ensure compliance and access control, protecting sensitive data and enhancing accountability [38]. These benefits align with management, operational, and business goals, although scalability issues and technical complexity limit broader adoption.

Decentralized Blockchain for Direct Transactions reduces costs, speeds up transactions, and builds trust with transparent ledgers, supporting management, operational, and business objectives [39]. Yet, technical complexity and scalability remain significant obstacles.

Smart Contracts and the Integration of Federated Learning with Blockchain for Credit Card Fraud Detection automate compliance and enhance privacy [40, 41]. These technologies holistically address management, operational, and business goals by boosting security, efficiency, and trust.

After Industry 4.0, Blockchain technology transforms the cybersecurity of banking sector by improving trust and efficiency, although challenges such as high costs, scalability, and integration issues hinder broader implementation.

*5.2.2.2 Analyse Blockchain Methods Used in Cybersecurity at Banks under Leavitt's Diamond Model*

Structurally, decentralized systems have minimized reliance on centralized control points, creating a more robust framework that can withstand disruptions and enhance overall stability.

Humans play the key role being when executing the consensus mechanism in blockchain but also have to compromise in privacy as their transactions could be evaluated and revealed to other users [35]. To effectively use blockchains, users need to adopt with new technology and fill their gaps in skills and knowledge.

With cryptographic methods, distributed ledgers, and federated learning, blockchain is a strong technology to streamline cross-border transactions, reduce fraud rate and save costs for banks and customers [37].

Blockchain has the capabilities of automating critical process at banks namely loan approvals and transaction settlements and identifying thefts and frauds. Especially, blockchain is speedy for the real-time transactions when compared with traditional banking systems to help banks gain the full control of their most up-to-date activities.

## 6  Cybersecurity at Banks in Policies and Practices

Drawing from the work of Mishra et al. (2022) [42], the chosen banks' policies and practice around cybersecurity are divided into three key areas: Online Banking & EPayment, Privacy and Identity Theft & Cybercrime. While these categories highlight the main areas of regulatory scrutiny that can be used to protect financial services organizations from cyber threats, there still remains a disconnect between policy and practice.

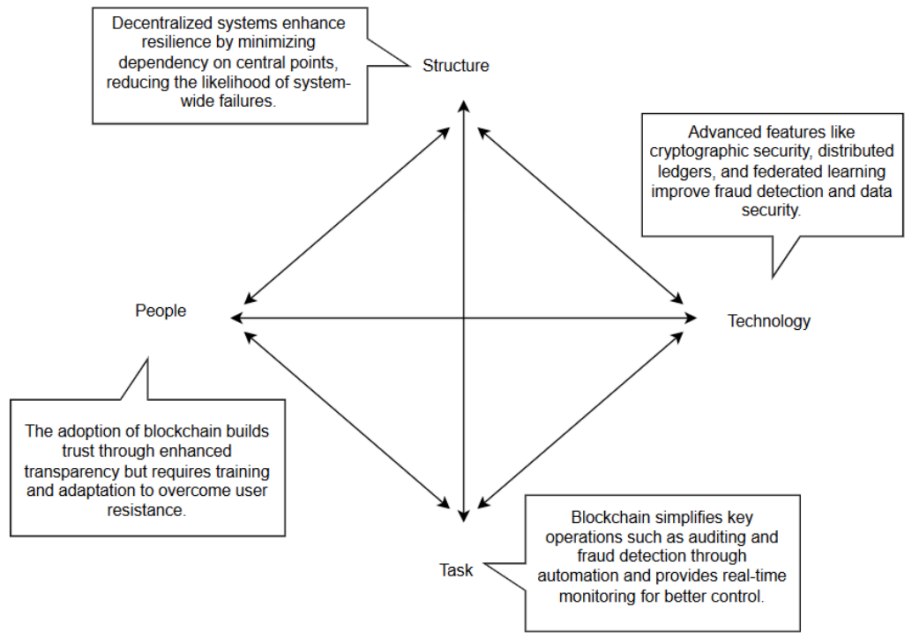

**Fig. 5.** Blockchain Cybersecurity Methods in Banking under Leavitt's Diamond model

### 6.1 Online Banking & E-Payment

For online banking security, the EU has strict policies such as the SWIFT network and Revised Payment Services Directive (PSD2). But the Bank of England said a worldwide payments problem affecting its CHAPS service, used for everything from large wholesale transactions to retail payments such as house purchases, had resulted in delays in some high-value and time-sensitive payments. Meanwhile, the European Central Bank said its settlement system had also been affected by the SWIFT outage [43]. The problem to some extent was resolved, but the SWIFT system needs, to some extent, long period lag between the application of SWIFT and the increase in speed of transaction. This approach of transaction finds a better solution, but they need to improve with more stability.

In the US, the Payment Card Industry Data Security Standard provides clear guidelines for safeguarding cardholder data. Yet the Verizon 2024 Data Breach Investigations Report (DBIR) found that financial services are still the most alluring of targets, with a significant number of breaches involving the use of stolen credentials [44]. Stolen credentials have been used in nearly a third (31%) of all breaches over the last decade, which really highlights that banks are still struggling to make their authentication systems strong.

## 6.2 Privacy

Despite the Cyber Privacy Fortification Act of 2015, the 2017 Equifax breach exposed 143 million Americans and thousands in the UK and Canada to identity theft. Equifax failed to patch a known vulnerability, revealing weaknesses in cybersecurity enforcement and compliance. This breach underscores how financial institutions often neglect regulatory safeguards, leaving consumers at risk [45].

The gap between policy and execution was further evident in Equifax's delayed response. Despite regulations requiring timely reporting, the company waited over a month before disclosing the breach, leaving consumers vulnerable. Their poorly managed response, including a flawed customer support website, further exposed users to phishing risks. This case highlights how financial institutions often fail to implement proactive cybersecurity measures and lack sufficient oversight, emphasizing the need for stricter enforcement and accountability.

## 6.3 Identity theft & Cybercrime

In attempts to thwart identity theft, new comprehensive policies have been established; however, their implementation by many banks has left them vulnerable to massive data breaches and financial loss.

For example, in 2022, the SEC charged JPMorgan Chase, UBS, and TradeStation for failing to comply with identity theft prevention requirements included in Regulation S-ID [46]. These banks lacked adequate mechanisms to identify and respond to identity theft red flags, compromising customers.

The most recent lawsuit, in 2024, by the Consumer Financial Protection Bureau, targeted large U.S. banks, such as Wells Fargo, Bank of America, and JPMorgan Chase, as well as the payment network Zelle, claiming that inadequate identity verification measures led to more than $870 million in fraud losses [47].

This suggests a misalignment of incentives, where financial institutions may focus more effectively on mitigating internal risks than preventing fraud against consumers, and raises systemic questions about consumer protection and how these cases are handled. These incidents highlight the policy-practice disconnect, with the need for more effective enforcement activity and proactive fraud detection in banks to meet regulatory intentions.

# 7 Discussion, Recommendation and Conclusion

## 7.1 Discussion

### 7.1.1 Artificial Intelligence

The strengths of AI in cybersecurity at banks include high accuracy in detecting intrusions, real-time management in the system and scalability for vast amount of data.

The automation of AI also replaces old manual processes in pre-industry 4.0, along with hybrid models when collaborating with other technologies such as blockchain and quantum cryptography.

Meanwhile, AI has dangers of overfitting, biases in training data, resource-intensive requirements and probability of being manipulated by adversarial inputs.

The future of AI is bright, when AI could handle big data, support banks in threat detection and predictive analytics and tailoring automating cybersecurity processes at banks. The widespreadness of open-source applications can grow the partnership between academia and industry for transformative applications

High-tech cybercriminals could take benefits from AI to deceive customers by using Deepfakes for pretending to be other people and building other complicated attacks that could be ignored by the system. The over-reliance on AI is risky without acknowledging the side effects of developing AI in the future.

### 7.1.2 Blockchain

With decentralized structure, cryptographic features, and unchangeable ledger provide, blockchain is a such a powerful framework for safeguard data and transactions, while minimising costs and upgrading efficiency. Along with state-of-the-art algorithms, blockchain could boosts the discovery and defense against cyber threats.

The real challenges of blockchain are expensive, scalability concerns, integration challenges and exact regulatory requirements to follow when adopting this technology.

In the future, hybrid model should be further developed with blockchain as it can collaborate with other innovative technologies such as Machine Learning, Deep Learning and Internet of Things to solve the fraud issues, create new services and refine financial payments.

The perils of leaking personal information, endpoint weaknesses, developing regulatory standards, and the competition with traditional banks must be tackled to unlock the full potential of blockchain.

## 7.2 Recommendation

### 7.2.1 *Artificial Intelligence*

Executives at banks need to regularly update the AI models for their companies to adapt with new attacks from cyberhackers and constantly adopt new methodology such as hybrid systems to increase the robustness of defence for the banks. Also, being thoughtful about the AI defence strategies and tailoring the education and training programs for staff could also be wise plans for the managers.

The investment in increasing the quality of data for AI training is critically necessary along with solving scalability and performance issues. Especially, executives need to be mindful about the biases when implementing AI such as gender, national and ethical biases to increase the equality and transparency of the system.

### 7.2.2 *Blockchain*

To leverage blockchain technology for stronger cybersecurity in banking, executives at banks should conduct regular security audits to help identify vulnerabilities and maintain system integrity. Comprehensive training for employees and stakeholders fosters deeper knowledge of blockchain, while user education encourages trust and smoother adoption.

Further investment in research and development is necessary to overcome technical challenges, improve system performance, and explore innovative solutions.

Partnering with blockchain experts and collaborating with regulatory bodies ensures legal compliance and the adoption of industry best practices. Establishing strong governance structures and security protocols minimizes risks, while exploring scalability options, such as sharding, enhances transaction processing capabilities. Pilot programs also allow banks to test blockchain applications in controlled settings, facilitating refinements before full implementation.

### 7.2.3  Practices

Banks can implement multi-factor authentication (MFA) and biometric authentication to improve security for their online banks and e-payments to reduce vulnerability from stolen credentials in breaches. By enhancing transaction monitoring systems with Artificial Intelligence-driven fraud detection mechanisms, it is possible to identify anomalies in real time. Moreover, payment processing infrastructure must always be updated by financial institutions to provide quicker and secure transactions while facing minimal downtime. Banks should also add warnings to the users while performing consecutive transactions to make them aware of potential fraudulent activity. Additionally, users would be expected to assist in limiting their banking experience by setting their maximum transaction amount.

Banks should routinely carry out security patching and vulnerability assessments for privacy protection to avoid breaches like Equifax. Establishing specialized incident response teams to effectively manage cybersecurity breaches as well as streams of clear communication to affected consumers will facilitate the response to security threats significantly.

In addition, educating customers about cybersecurity best practices going forward, like phishing detection and how to avoid identity theft, will help to reduce consumer vulnerabilities. Banks and financial institutions can use AI-based identity verification solutions that help in preventing identity theft and cybercrime by detecting fraud before the transactions get completed. This will make internal risk assessments for fraud more robust and allow them to act on risk and minimize fraud proactively. Moreover, banks need to implement continuing live surveillance of customer accounts for suspect transactions and implement customer alerts if unusual transaction behavior occurs, which would help to enhance consumer trust and consumer protection.

These practices can also help banks to better adopt and enforce cybersecurity policies, thereby connecting regulatory frameworks with real-world security measures.

## 7.3 Conclusion

This paper explored the evolution of cybersecurity in the banking sector, highlighting the transition from pre-industry 4.0's traditional methods to the advanced technologies of Artificial Intelligence and blockchain in post-industry 4.0. By adopting these innovations and fostering collaboration, banks can effectively address modern cyber threats while maintaining security and trust in the digital age.

## Acknowledgement

Thank you, Dad (Mr. Tran Xuan Tuyen) and Mom (Mrs. Le Thi Phuong), for always supporting me unconditionally. I'm grateful for having my Aunt (Ambassador Le Thi Tuyet Mai), whom I consider my second Mom, to be my inspiration.

I appreciate the guidance from Dr. Kushwanth Koya at The University of Sheffield at the beginning to lead to this paper.

## Declarations

- Funding: Not applicable.
- Competing interests: Not applicable.
- Data availability: The data used in this study were obtained from publicly available academic databases including Google Scholar, IEEE Xplore, Scopus, Springer and ACM. No confidential data was used.
- Author contribution: The sole author conducted all aspects of this research, including conceptualization, data collection, analysis, writing and review.

[14] Karim, Z., Rezaul, K. M., & Hossain, A. (2009). Towards secure information systems in online banking. In 2009 International Conference for Internet Technology and Secured Transactions (ICITST). https://doi.org/10.1109/icitst.2009.5402619.

[15] Mahdi, M. D. H., Rezaul, K. M., & Rahman, M. A. (2010, February). Credit fraud detection in the banking sector in UK: a focus on e-business. In 2010 Fourth International Conference on Digital Society (pp. 232–237). IEEE.

[16] M¨ockel, C., & Abdallah, A. E. (2010, August). Threat modeling approaches and tools for securing architectural designs of an e-banking application. In 2010 Sixth International Conference on Information Assurance and Security (pp. 149–154). IEEE.

[17] He, K., Feng, Z., & Li, X. (2008, December). An attack scenario-based approach for software security testing at the design stage. In 2008 International Symposium on Computer Science and Computational Technology (Vol. 1) (pp. 782–787). IEEE.

[18] McCombie, S., & Pieprzyk, J. (2010, July). Winning the phishing war: A strategy for Australia. In 2010 Second Cybercrime and Trustworthy Computing Workshop (pp. 79–86). IEEE.

[19] Cha, B., & Park, S. (2008, November). Design and Efficiency Analysis of New OTP System Using Homomorphic Graph of Fingerprint Features. In 2008 Third International Conference on Convergence and Hybrid Information Technology (Vol. 2) (pp. 585–590). IEEE.

[20] de la Puente, F., Sandoval, J. D., & Hernandez, P. (2003, October). Pocket device for authentication and data integrity on Internet banking applications. In IEEE 37th Annual 2003 International Carnahan Conference on Security Technology (pp. 43–50). IEEE.

[21] Deng, P. S., Jaw, L. J., Wang, J. H., & Tung, C. T. (2003, October). Trace copy forgery detection for handwritten signature verification. In IEEE 37th Annual 2003 International Carnahan Conference on Security Technology, 2003. Proceedings. (pp. 450-455). IEEE.

[22] Podebrad, I., & Drotleff, M. (2009, May). IT Security in Banking - Processes, Practical Experiences, and Lessons Learned. In 2009 Fourth International Conference on Internet Monitoring and Protection (pp. 78–83). IEEE.

[23] Hu, X., Zhao, G., & Xu, G. (2009, January). Security scheme for online banking based on secret key encryption. In 2009 Second International Workshop on Knowledge Discovery and Data Mining (pp. 636–639). IEEE.

[24] Zihao, S. (2010). An improved SET protocol payment system. In 2010 International Conference on Computer and Communication Technologies in Agriculture Engineering.

[25] Gandotra, V., Singhal, A., & Bedi, P. (2009, October). Threat mitigation, monitoring and management plan-a new approach in risk management. In *2009 International Conference on Advances in Recent Technologies in Communication and Computing* (pp. 719-723). IEEE.
26